\documentstyle[psfig]{article}
\font\t=cmbx10 at 16pt
\font\au=cmr10 
\font\ad=cmti10 at 9pt 
\font\s=cmbx10 
\font\sb=cmti10 
\font\fc=cmr10 at 9pt 
\begin{document}
\begin{center}
{\t High Resolution Imaging by Employing Passive and Active Approaches} 
\end{center}
\vspace{0.7cm}

\begin{center}
{\au Swapan K. Saha}\\                           
{\ad Indian Institute of Astrophysics,
Bangalore - 560 034, India}
\end{center}
\vspace{1cm}

{\au
\noindent
{\s 1. Introduction}
\\

\noindent
The Earth's atmosphere is a highly turbulent medium that 
distorts the characteristics of downward propagating light; the
longer the path, the more it suffers deflection. Light reaching the 
entrance pupil of an imaging system is coherent only within 
patches of diameter, $\approx r_\circ -$ Fried's
parameter [1]. This causes blurring of the image which limits
the theoretical performance of any terrestrial large optical telescope as 
against lone orbiting one. The blurring suffered by such an image is
modeled by a convolution with a point spread function (PSF). 

The diffraction limited resolution of celestial objects viewed through such
turbulence could be achieved by employing post-detection processing of a large 
data set of short exposure images using Fourier domain methods. Certain 
specialized moments of the Fourier transform of a short exposure image contain 
diffraction limited information about the object of interest. In a classic 
paper, Labeyrie [2] suggested a method called, speckle interferometry, 
where post-detection data processing algorithms are required to decipher 
diffraction limited spatial Fourier spectrum and image features of stellar 
objects. Though it is a passive approach, this method has opened up a new era 
in modern optics and has made impacts in several important fields in 
astrophysics [3, 4], viz., (i) in studying the separation 
and orientation of close binary stars [5], (ii) in measuring diameter of giant 
stars [6, 7], (iii) in resolving Pluto-Charon system [8], (iv) in determining 
shapes of asteroids [9], (v) in mapping the finer features of extended objects 
[10], (vi) in estimating sizes and mapping certain types of circumstellar 
envelopes [11, 12], (vii) in revealing structures of active galactic nuclei 
[13], and of compact clusters of a few stars like R136a complex [14], (viii) in 
resolving the gravitationally lensed QSO's [15], etc. 

Following its success, astronomers focussed their efforts on developing
post-detection image processing techniques and applied them to improve the
resolution of astronomical images [3, 4]. 
Considerable amount of new informations have also come in 
by employing various other avenues at the telescopes, namely, (i) speckle
spectroscopy, (ii) speckle polarimetry, (iii) shear interferometry, (iv)
phase closure method, (v) aperture synthesis technique, (vi) differential
speckle interferometry, (vii) phase diversity imaging etc.
An interesting result from the speckle spectroscopic observations of a
binary system, $\phi$~And by Baba et al., 
[16] is that the primary star (Be star) has an H$\alpha$ emission line while 
the companion has an H$\alpha$ absorption line.

Significant improvements in technological innovation over the past several years
have brought the hardware to compensate perturbations in the wavefronts in real 
time [17, 18]. This active approach, known as adaptive optics (AO) system has 
advantages over a passive approach that are 
limited by noise. It is able to recover near diffraction limited images and 
improves the point source sensitivity. One of its most successful applications
in astronomy has been in imaging of Neptune's ring arcs. AO systems
are employed in other branches of physics as well.
Liang et al. [19] have constructed a camera equipped with adaptive-optics that
allows one to image a microscopic size of single cell in the living human
retina. They have shown that a human eye with adaptive-optics correction can
resolve fine gratings that are invisible to the unaided eye. AO systems are
useful for spectroscopic observations, as well as for photon-starved imaging
with future very large telescopes, and ground based long baseline optical
interferometers [20, 21].

In depth study has been made by Sirohi [22-24] on speckle metrology dealing with
the coherent monochromatic source for over a 
period of two decades, and a few glimpses of such exemplary works can be 
witnessed in his recently published article [24]. Since I am familiar with the
subject [25] that he was working on, I begin the paper with the  
discussions on the formation of speckles in the case of non-coherent quasi
monochromatic source and of ways to detect them. Prior to this, a brief
introduction on the behavior of the atmosphere and its effect on the plane 
wavefront from a stellar source is presented. The benefit of short-exposure 
images over long exposure is highlighted with emphasis set on their comparison 
as well. The data processing methods that are implemented in our 
institute in order to decipher both Fourier modulus and Fourier phase are 
described. The salient features of the adaptive optics system are discussed. 
At last but not the least, I conclude the paper after
a brief discussion on certain astrophysical problems that can be targeted using 
moderate sized telescopes available in India, such as 2.34 meter
Vainu Bappu Telescope (VBT), Kavalur, 2 meter Himalayan Chandra Telescope
(HCT), Hanley. 
\\

\noindent
{\s 2. Atmospheric turbulence and Seeing}
\\

\noindent
Due to the turbulence and the concomitant development of thermal convection in 
the atmosphere, the density of air fluctuates in space and time. 
According to the Kolmogorov's theory of fluid turbulence [26], when 
the Reynolds number exceeds some value in a pipe (depending on the geometry of
the pipe), the transition from laminar flows to turbulent flows occurs. The 
dimensionless quantity Reynolds number is defined as ${\it v}_0L/\nu$, where 
${\it v}_0$ is the 
mean flow speed, $L$ is the transverse size of the pipe and $\nu$ is kinematic
viscosity of the fluid. If $L$ is taken as some characteristic size of the
flow of atmosphere the result holds good for atmospheric case. When the kinetic
energy of the air motions at a given length scale is larger than the energy
dissipated as heat by viscosity of the air at the same scale - the kinetic
energy of large scale motions would be transferred to smaller and smaller
motions; motions at small scales would be statistically isotropic at the small
scales, viscous dissipation would dominate the breakup process.
The distribution of turbule sizes ranges from millimeters 
to meters, with lifetimes varying from milliseconds to seconds. 

Heating of the Earth's
atmosphere by solar radiation causes turbulent motions in the atmosphere.
During day time, large warm packets of air closer to the ground move up
due to buoyancy and initiate convection causing the turbulence near the
ground. They dissipate their kinetic energy continuously and randomly into
smaller and smaller packets of air, each having a unique temperature. These
packets are called turbulent eddies. Convection changes with insolation
and disappears during night time. However, horizontal circulation of air starts.
An important property of eddies is that they exist in a variety of
length scales and their distribution is random. There exists an upper limit,
$L_0$, decided by the process that generates turbulence and a lower limit,
$l_0$, decided by the size at which viscous dissipation overtakes the breakup
process.

Turbulent air motions cause variations in the density,
pressure, temperature and humidity of the air from one point to another.
While the local temperature fluctuations of the air could be of the order of
few hundredth of a degree throughout the atmosphere, fluctuations of a few
tenths of a degree or more are typical in the lowest layer of the atmosphere.
The temperature fluctuations in small patches of air cause changes in index of 
refraction. The power spectral density of these refractive index fluctuations 
follows a power law with large eddies having greater power [27].   
The refractive index of the atmosphere may be expressed as, 
\begin{equation}
n({\bf r, t}) = n_0 + n_1({\bf r, t}), 
\end{equation}

\noindent
where, $n_0 \approx 1$ is the mean refractive
index of air, $n_1({\bf r, t})$ the randomly fluctuating term, ${\bf r}$ 
the 3-dimensional (3-D) position vector and $t$ the time. The refractive 
index varies due to the temperature inhomogeneities. The dependence of the 
refractive index of air
upon pressure, $P$ (millibar) and temperature, $T$ (Kelvin), at optical
wavelengths is given by [28],
\begin{equation}
n_1 \cong n - 1 = 77.6~\times~10^{-6}\frac{P}{T}.
\end{equation}

\noindent
As the refractive index of the air is highly sensitive to the temperature,
it varies randomly from one point to another. Fluctuations in the
refractive index induce random optical path lengths to the ray that are normal
to the wavefront arriving at the top of the atmosphere from a distant star. 
Consequently,
the wavefront reaches the external pupil of a ground based telescope, 
gets corrupted in the sense that the surface of constant phase is no longer 
planar; it has an overall tilt and small scale corrugations on top of it. The 
RMS value of the phase perturbations increases with the size of the wavefront.

The perturbations in the wavefront produce effects similar to optical 
aberrations in the telescope and thus degrade the image quality. When a very 
small aperture is used, a small portion of the wavefront is intercepted and 
the phase of the wavefront is uniform over the aperture. If the amplitudes of 
the small scale corrugations of the wavefront are much smaller than the 
wavelength of the light, the instantaneous image of a star is sharp  
resembling to the classical diffraction pattern taken through an ideal 
telescope, in which the point spread function (PSF) is invariant to 
spatial shifts. In the absence of the atmospheric turbulence,
FWHM, known as Airy disk, is the diffraction limit of telescope, i.e., 
$\theta \sim \lambda/D$, where $\lambda$ is a wavelength of light and $D$ the 
diameter of the telescope. 

The discrete layers of turbulence are blown by wind across the telescope 
aperture, hence a change in tilt occurs, which in turn,
causes random motion of the star's image at the focal plane. As the aperture
size increases, there is a decrease in the sharpness and amplitude of the
motion; the amplitude of the random variation of phase across the intercepted 
wavefront tends to become larger. This leads to the blurring of the image, 
which is larger than the Airy disk of the telescope. 
For a 2 meter telescope, the size of this Airy disk in the absence of the
atmospheric turbulence, is 0.05~arcsec~($^{\prime\prime}$) at 0.5$\mu$m. 
While in the 
presence of such turbulence the size of the image becomes typically 
0.5 - 2$^{\prime\prime}$. The resolution at the image plane of a ground based 
telescope is dictated by the width of the PSF of both the atmosphere and the 
telescope, which is of the order of $(1.22\lambda/r_0)$, where  
$r_0$ is the average size of the turbulence cell [1]. For a good
site, the typical value of $r_0 \sim$ 20~cm, therefore, any telescope larger
than $r_0$ cannot provide better spatial resolution. 
The image motion and blurring together are referred to as
atmospheric seeing or simply seeing.
\begin{figure}
\centerline{\psfig{figure=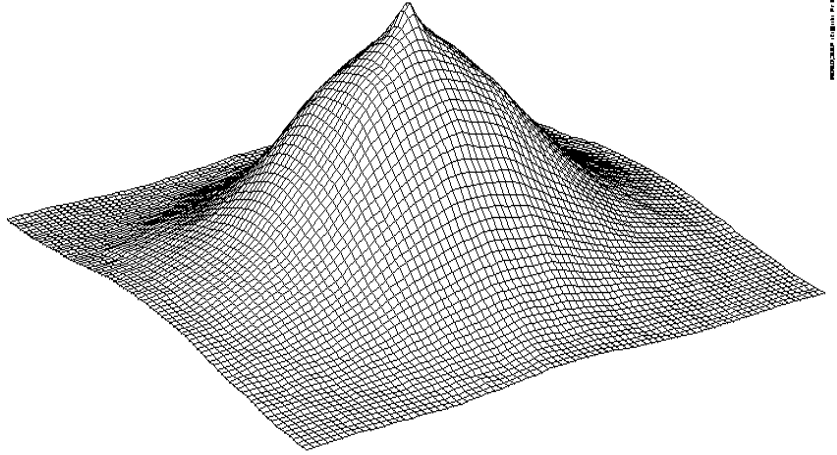,height=8.0cm,angle=270}}
\caption{\fc 3-D picture of the autocorrelation of HR~2305 observed at 
2.34~m VBT, Kavalur, India, on 28 March 1991, at 1510 UT [29].}
\end{figure}

Seeing is the total effect of distortion in the path of the light via different 
contributing layers of the atmosphere to the detector placed at the focus of 
the telescope [3, 4]. The major sources of image 
degradation predominantly comes from the surface layer, as well as from 
the aero dynamical disturbances in the atmosphere surrounding the 
telescope and its enclosure, namely, (i) thermal distortion of primary and 
secondary mirrors when they get heated up, (ii) dissipation of heat by the
latter mirror, (iii) rise in temperature at the primary cell, (iv) at the focal
point causing temperature gradient close to the detector etc. Degradation in
image quality can occur due to opto-mechanical aberrations such as aberrations 
in the design, manufacture and alignment of the optical train, mechanical
vibrations of optical system as well.

Saha and Chinnappan [29] have computed the night time variations of  
$r_0$ that varies at random at the 2.34 meter Vainu 
Bappu Telescope (VBT) site, Kavalur, India using the data obtained by means of
speckle imaging. The averaged autocorrelation of the these images contains 
both the autocorrelations of the seeing disk, as well as of the
mean speckle cell. It is the width of the speckle component of the
autocorrelation that provides the information on the size of the object being
observed. The form of transfer function, 
$<\mid\widehat{S}({\bf u})\mid^2>$, is obtained by calculating
Wiener spectrum of the instantaneous intensity distribution from a point
source. Figure 1 depicts the autocorrelation of HR~2305 observed at 1510 hrs UT,
obtained at the Cassegrain focus of 2.34~m VBT, Kavalur, India [29].
They found that the average observed $r_\circ$ is
higher during the later part of the night than the earlier part, implying that
the PSF has a smaller FWHM during the former period. This might indicate that
the slowly cooling mirror creates thermal instabilities that decreases slowly
over the night. Figure 2 depicts the night time variations of $r_0$ at 
2.34 meter
VBT, Kavalur [29].
\begin{figure}
\centerline{\psfig{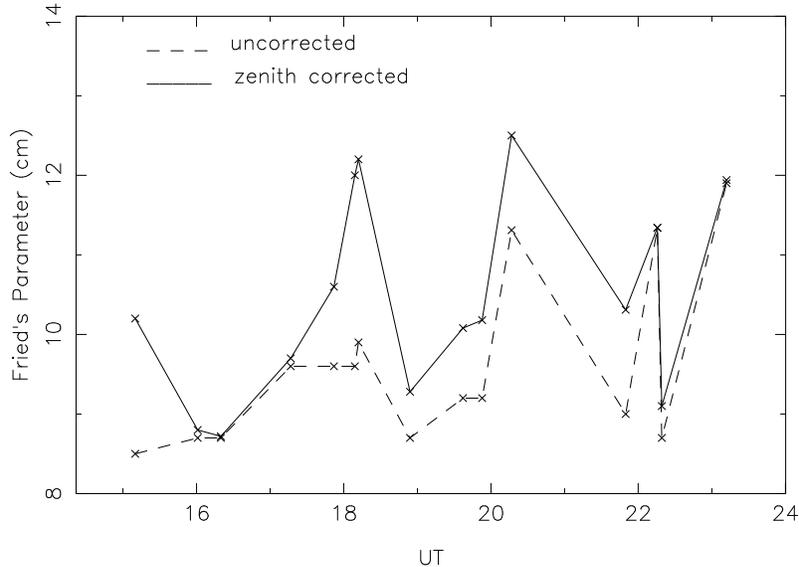}}
\caption{\fc Nighttime variations of $r_0$ at the 2.34 meter VBT, VBO, Kavalur
[29].}
\end{figure}
\\

\noindent
{\bf 3. Speckle imaging}
\\

\noindent
Ever since the development of the speckle interferometric technique by Labeyrie
[2], it is widely used both in the visible, as
well as in the infrared (IR) bands. The description of speckle interferometry,
runs as follows.
\\

\noindent
{\sb 3.1 Speckle formation and statistics}
\\

\noindent
The term, speckle, refers to the grainy structure observed when an uneven
surface of an object is illuminated by a fairly coherent source. The formation
of speckles stems from summation of coherent vibrations having different random
characteristics. The statistical properties of speckle pattern depend on both
the coherence of the incident light, as well as on the random properties
of the medium. Depending on the randomness of the source, spatial or temporal
speckles tend to appear; spatial speckles may be observed when all parts of the
source vibrate at same constant frequency but with different amplitude and
phase, while the latter are produced if all parts have uniform amplitude
and phase. With a heterochromatic vibration spectrum, in the case of random
sources of light, spatio temporal speckles are produced. Their statistical
properties depend both on the coherence of the incident light and the random
properties of medium. Since the positive and negative values cannot cancel out
everywhere, adding an infinite number of such sine functions would result in a
function with 100$\%$ constructed oscillations [30].

If a point source is imaged through
the telescope by using the pupil function consisting of two subapertures
$(\theta_1, \theta_2)$ corresponding to the two seeing cells separated by a
vector $\lambda u$, a fringe pattern is produced with narrow spatial frequency
bandwidth that moves within broad PSF envelopes with increasing distance between
the subapertures, the fringes move with an increasingly larger amplitudes. The
introduction of third aperture provides three pairs of subapertures and yields
the appearance of three intersecting patterns of moving fringes. With $r_0$ 
sized subapertures covering the telescope aperture synthesizes a filled aperture
$p_j$ (each pair of them, $p_n, p_m$ separated by a baseline interferometer). 
The intensity at the focal plane $I$, according to the theory of diffraction 
[31] is determined by the expression,
\begin{equation}
I= \sum_{n,m} \left<\Psi_n \Psi^{*}_m\right>,
\end{equation}

\noindent
The term $\Psi_n \Psi^{*}_m$, is multiplied by $e^{i\psi}$, where $\psi$ is the
random instantaneous shift in the fringe pattern. Each sub aperture is small
enough for the field to be coherent over its extent. Atmospheric turbulence
does not affect the contrast of the fringes but their phases are randomly 
distorted. If the exposure time is shorter than the evolution time of the phase
inhomogeneities, then each patch of the wavefront with diameter $r_0$
would act independently of the rest of the wavefront
resulting in multiple images of the source. These sub images or speckles, as 
they are called and spread over the area defined by the long exposure image,
can occur randomly along any direction within an angular patch of diameter 
$\lambda/r_0$. The average size of the speckle is of the same order of 
magnitude as the Airy disk of the telescope in the absence of atmospheric 
turbulence. 
\begin{figure}[h]
\centerline{\psfig{figure=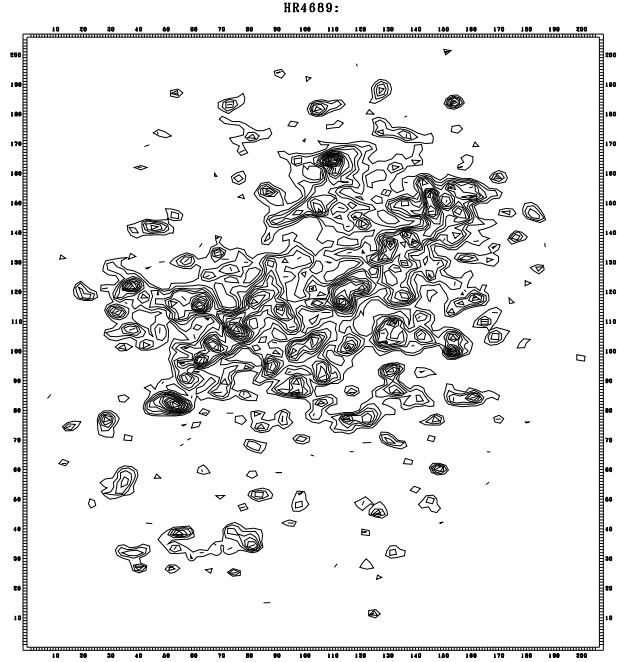,height=9.5cm,angle=270}}
\caption{\fc Specklegram of a binary star, HR~4689 obtained at VBT, Kavalur, 
India.  }
\end{figure}

A specklegram of a point source is composed of numerous shortlived speckles.
Constructive interference of the fringes would show an enhanced bright speckle. 
The number of correlation cells is proportional to the square of $D/r_0$, and 
the number of photons, $N_p$, per speckle, but is independent of its diameter. 
The lifetime of speckles is of the order of 0.1 to 0.01 seconds, and is
determined by $\tau_0 \approx r_0/\Delta{\it v}$, in which $\Delta{\it v}$ is 
the velocity dispersion in the turbulent seeing layers across the line of sight.
Integration time of each exposure varies from a
few milliseconds to twenty milliseconds, depending on the condition of seeing,
to freeze the single realization of the turbulence. 
\\

\noindent
{\sb 3.2. Conventional image}
\\

\noindent
The variance of phase difference fluctuations between any two points in the 
wavefront increases as the 5/3 power of their separation. When this variance
exceeds $\pi^2$~rad for some separation $r_0$, then all details 
smaller than $\lambda/r_0$ will be obliterated in the long exposure images. 

Figure 3 depicts the speckles of the star
HR~4689; observations were carried out at 2.34 meter VBT,
Kavalur, India with the speckle interferometer [32, 33]. 
A snap shot taken later would depict a different pattern but with similar 
probability of the angular distribution. A sum of similar exposures is the
conventional image. It is easy to visualize that the sum of several 
statistically uncorrelated speckle patterns from a point source can result in 
an uniform patch of light a few arcseconds wide [34]. 
Figure 4 shows the result of summing 128 specklegrams demonstrating the 
destructions of finer details of the image by the atmospheric turbulence.
\begin{figure}[h]
\centerline{\psfig{figure=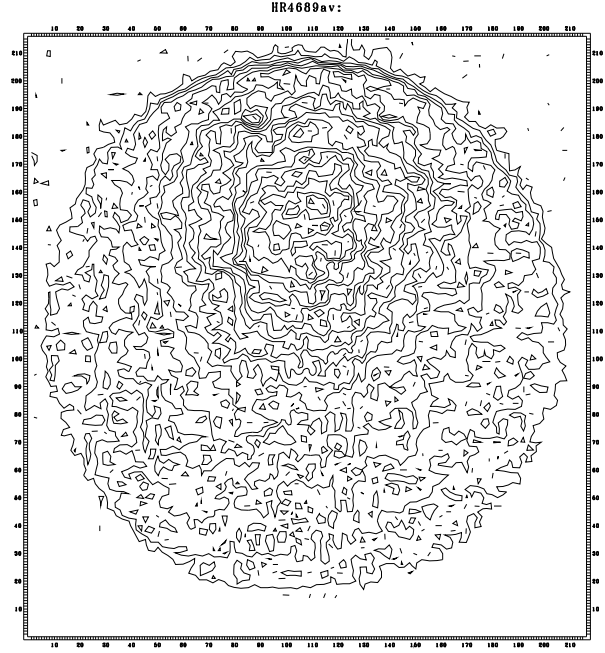,height=9.5cm,angle=270}}
\caption{\fc The result of summing 128 specklegrams of the same star, HR~4689. 
}
\end{figure}

The intensity distribution of the image, $I({\bf x})$, is the convolution
of intensity distribution of the object $O({\bf x})$ and the 
PSF, $S({\bf x})$, i.e.,
\begin{eqnarray}
I({\bf x}) &=& \int O({\bf x}^\prime) S({\bf x} -
{\bf x}^\prime) d{\bf x}^\prime \nonumber\\
&=& O({\bf x}) \star S({\bf x}),
\end{eqnarray}

\noindent
where ${\bf x} = (x, y)$ is two dimensional (2-D) space vector, and $\star$ 
stands for the convolution. The Fourier transform of this intensity 
$I({\bf x})$ is defined by,
\begin{equation}
\widehat{I}({\bf u}) = \int_{-\infty}^\infty I({\bf x})e^{-i2\pi{\bf u}.{\bf x}}
d{\bf x},
\end{equation}

\noindent
where ${\bf u} = (u, v)$ the 2-D spatial frequency vector. 
In the case of the long exposure, the PSF is defined by its ensemble average, 
$\left<S({\bf x})\right>$, and therefore, the average 
illumination, $\left<I({\bf x})\right>$ is given by, 
\begin{equation}
\left<I({\bf x})\right> = O({\bf x}) \star \left<S({\bf x})\right>,
\end{equation}

\noindent
where $\left< \right>$ indicates the ensemble average. 
By using 2-D Fourier transform (FT), this equation can be read as,
\begin{equation}
\left<\widehat{I}({\bf u})\right> = \widehat{O}({\bf u}) \cdot \left<\widehat{S}
({\bf u})\right>,
\end{equation}

\noindent
where $\widehat{O}({\bf u})$ is
the object spectrum, $\left<\widehat{S}({\bf u})\right>$ the transfer function 
of the atmosphere and the telescope for long exposure images.
\\

\noindent
{\sb 3.3 Speckle interferometry}
\\

\noindent
Speckle interferometry [2] estimates a power spectrum which is an ensemble
average of the squared modulus of an ensemble of FT from a
set of specklegrams that represent the resultant of
diffraction-limited incoherent imaging of the object irradiance
convolved with the function representing the combined
effects of the turbulent atmosphere and the image forming optical system.
An ensemble of such specklegrams, $I_n({\bf x}),
n = t_1, t_2, t_3, \ldots, t_N$, constitute an astronomical speckle observation.
By integrating the autocorrelation function of these successive narrow bandpass
exposures, the spatial resolution of the objects at low light levels can be
obtained. The transfer function of $S({\bf x})$, can be estimated by
calculating Wiener spectrum of the instantaneous intensity from the unresolved
star. The size of the data sets is constrained by the consideration of the
signal-to-noise (S/N) ratio. Usually specklegrams of the
brightest possible reference star are recorded to ensure that the S/N ratio of
reference star is much higher than the S/N ratio of the programme star.

The variability of the corrugated wavefront yields speckle
boiling and is the source of speckle noise that arises from difference in
registration between the evolving speckle pattern and the boundary of the PSF
area in the focal plane. These specklegrams have additive noise contamination,
$N_j({\bf x})$, which includes all additive measurement of uncertainties.
This may be in the form of (i) photon statistics noise, and (ii) all distortions
from the idealized isoplanatic model represented by the convolution of
$O({\bf x})$ with $S({\bf x})$ that includes nonlinear geometrical distortions. 
Due to (i) variations of air mass or of its
time average between the object and the reference, (ii) differences in seeing
between the MTF for the object and its estimation from the reference (such a
comparison is likely to introduce deviation in the statistics of speckles from
the expected model based on the physics of the atmosphere), (iii) deformation
of mirrors or misalignment while changing its pointing direction,
(iv) bad focusing, (v) thermal effect from the telescope etc., the quality of
the image degrades, which, in turn, would result either in
the suppression or in the enhancement of intermediate spatial frequencies. This
may lead to a dangerous artifact, yielding a wrong identification of the
companion star [35]. Therefore, it is essential to choose the point
source calibrator as close as possible for which all observing conditions are
required to be identical to those for the object, preferably within 1$^\circ$
of the program star. The object and calibrator observations should be
interleaved to calibrate for changing seeing condition by shifting the
telescope back and forth during the observing run to equalize seeing
distributions. Another difficulty arises from the non-detectability of
a pair of photons closer than a minimum separation by the frame transfer CCD
that is subjected to limitations in recording fast photon-event pairs, yielding
a loss in high frequency information. This, in turn, produces a hole in the 
center of the autocorrelation $-$ Centreur hole, resulting in the degradation 
of the power spectra or bispectra (Fourier transform of the triple
correlation) of short exposures images.

For each of the short exposure instantaneous records, the instantaneous image 
intensity, $I({\bf x})$ is given by,
\begin{equation}
I({\bf x}) = O({\bf x}) \star S({\bf x}) + N({\bf x}),
\end{equation}

\noindent
In the Fourier plane, the effect becomes a multiplication, point by point, of
the transform of the object, $\widehat{O}({\bf u})$ with the transfer function,
$\widehat{S}({\bf u})$, thus,
\begin{equation}
\widehat{I}({\bf u}) = \widehat{O}({\bf u}) \cdot \widehat{S}({\bf u}) + 
\widehat{N}({\bf u}).
\end{equation}

\noindent
The ensemble average of the power spectrum is given by,
\begin{equation}
\left<\left|\widehat{I}({\bf u})\right|^{2}\right> = \left|\widehat{O}({\bf u})
\right|^{2} \cdot \left<\left|\widehat{S}({\bf u})\right|^{2}\right> + 
\left<\left|\widehat{N}({\bf u})\right|^2\right>.
\end{equation}

\begin{figure}[h]
\centerline{\psfig{figure=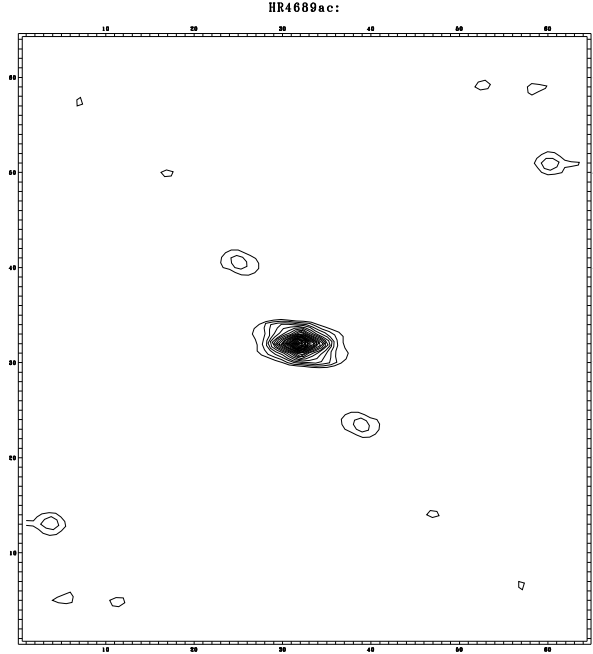,height=9.5cm,angle=270}}
\caption{\fc Autocorrelation of a binary star, HR~4689 [36]; the companion star 
is one of the contours that are symmetrically placed on either side of the 
main peak. The contours at the edge are the artifacts.}
\end{figure}

Saha and Maitra [36] developed an algorithm, where a Wiener parameter,
$w_1$, is added to PSF power spectrum which helps reconstruction with a few
realizations.
\begin{equation}
\left|\widehat{O}{\bf (u)}\right|^2 = \frac{\left<\left|\widehat{I}{\bf (u)}
\right|^2\right>}{\left[\left<\left|\widehat{S}{\bf (u)}\right|^2\right> + 
w_1\right]}.
\end{equation}

\noindent
The speckle interferometry in the case of the components in a
group of stars retrieves the separation, position angle with 180$^\circ$
ambiguity, and the relative magnitude difference at low light levels.
Figure 5 depicts the autocorrelation of a binary system, HR~4689.
\begin{figure}[h]
\centerline{\psfig{figure=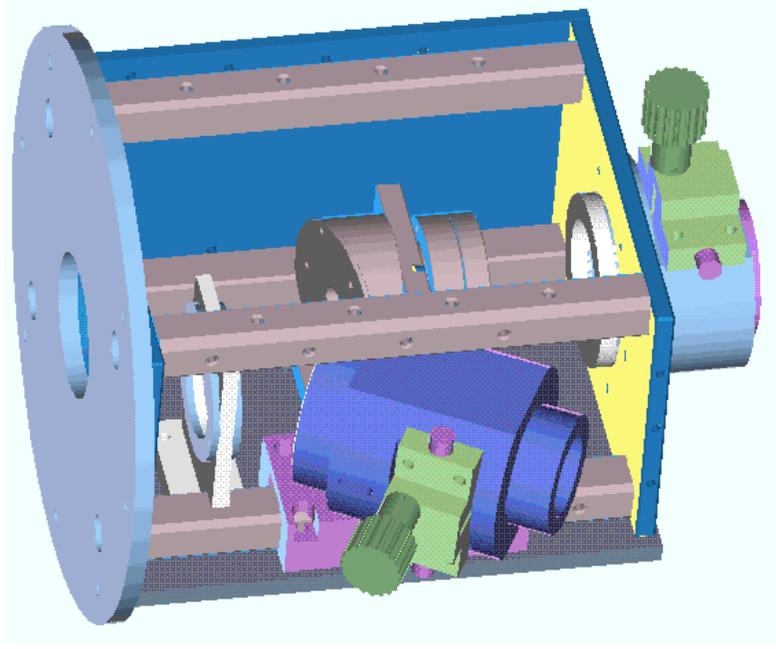,height=8.5cm,angle=270}}
\caption{\fc Overview of the Speckle camera system that is developed by
Saha et al., [32, 33] for the use at VBT, Kavalur, India.}
\end{figure}

The programme of observing close binary systems (separation $<1^{\prime\prime}$)
has been going on since 1996 using a speckle camera system (see Figure 6) at 
the Cassegrain focus of the 2.34 VBT, Kavalur, India [32, 33]. 
This camera system samples the image scale at the Cassegrain focus of the 
said telescope to 0.015$^{\prime\prime}$ per pixel of the intensified CCD. The 
wavefront falls on the focal plane
and passes on to a microscope objective through a circular aperture of 
$\sim$350~$\mu$m of an optical flat kept at an angle of 15$^\circ$.  
The enlarged beam is recorded after passing through a
narrow band filter by a Peltier cooled based solid state electron multiplying 
CCD based camera. Unlike the uncooled ICCD where data is stored in 8 bits, 
in this system, data is stored to 16 bits and can be archived to a Pentium PC. 
The surrounding star field of diameter $\sim$~10~mm gets
reflected from the optical flat on to a plane mirror and is reimaged on to 
an uncooled ICCD [37] for guiding the object.
\\

\noindent
{\s 4. Deciphering Fourier phase}      
\\

\noindent
Several algorithms have been developed to retrieve the diffraction limited 
phase of a degraded image, of which the triple correlation technique, known as 
speckle masking method [38, 39], is used by 
many observers. It is a generalization of closure phase technique where the 
number of closure phases is small compared to those available from bispectrum. 
The notable advantage of such an algorithm is that
it is insensitive to (i) the atmospherically induced random phase 
errors, (ii) the random motion of the image centroid, and (iii) the permanent 
phase errors introduced by telescope aberrations; any linear phase term in the 
object phase cancels out as well. The images are not required to be shifted to 
common centroid prior to computing the bispectrum. The other advantages are: 
(i) it provides information about the object phases with better S/N ratio from 
a limited number of frames, and (ii) it serves as the means of image recovery 
with diluted coherent arrays [40]. The disadvantage of this technique is that 
it demands severe constraints on the computing facilities with 2-D data since 
the calculations are four dimensional (4-D). It requires 
extensive evaluation time and data storage requirements,
if the correlations are performed by using digitized images on a computer.

A triple correlation is obtained by multiplying a shifted object, 
$I({\bf x} + {\bf x}_1)$ is multiplied with the original object, $I({\bf x})$,
followed by cross correlating the product mask, 
$I({\bf x})I({\bf x}+{\bf x}_1)$ with the original one.  
The triple correlation of a specklegram is given by,
\begin{equation}
I({\bf x}_1,{\bf x}_2) = \int_{-\infty}^\infty I({\bf x})
I({\bf x} + {\bf x}_1) I({\bf x} + {\bf x}_2)d{\bf x},
\end{equation}

\noindent
where ${\bf x}_j = {\bf x}_{jx} + {\bf x}_{jy}$ are 2-D spatial coordinate
vectors. The Fourier transform of the triple correlation is known as 
bispectrum. The ensemble averaged bispectrum is expressed as,
\begin{eqnarray}
\widehat{I}({\bf u}_1,{\bf u}_2) &=& \left<\widehat{I}({\bf u}_1)
\widehat{I}^\ast({\bf u}_1 + {\bf u}_2) \widehat{I}({\bf u}_2)\right>,\\
&=& \widehat{O}({\bf u}_1)\widehat{O}^\ast({\bf u}_1 + {\bf u}_2) 
\widehat{O}({\bf u}_2) \left<\widehat{S}({\bf u}_1)
\widehat{S}^\ast({\bf u}_1 + {\bf u}_2) \widehat{S}({\bf u}_2)\right>,
\end{eqnarray}

\noindent
where ${\bf u}_j = {\bf u}_{jx} + {\bf x}_{jy};$ $\widehat{I}({\bf u}_j)$
and $\widehat{I}^\ast({\bf u}_1 + {\bf u}_2)$ denote Fourier transforms of
$I({\bf x})$, i.e., 
$\widehat{I}({\bf u}_j) = \int_{-\infty}^\infty I({\bf x})e^{-i2\pi{\bf u}_j.
{\bf x}}d{\bf x}$, and 
$\widehat{I}^\ast({\bf u}_1 + {\bf u}_2) = \int_{-\infty}^\infty I({\bf x})
e^{-i2\pi({\bf u}_1 + {\bf u}_2).{\bf x}}d{\bf x}$. 
The object bispectrum is given by, 
\begin{equation}
\widehat{O}({\bf u}_1,{\bf u}_2) = \widehat{O}({\bf u}_1)
\widehat{O}^\ast({\bf u}_1 + {\bf u}_2) \widehat{O}({\bf u}_2) 
= \frac{\left<\widehat{I}({\bf u}_1)
\widehat{I}^\ast({\bf u}_1 + {\bf u}_2) \widehat{I}({\bf u}_2)\right>}{
\left<\widehat{S}({\bf u}_1)
\widehat{S}^\ast({\bf u}_1 + {\bf u}_2) \widehat{S}({\bf u}_2)\right>}.
\end{equation}

The bispectrum of a 2-D intensity distribution is a 4-D function. Due to 
this extension in to 4-D space it is possible that the phase information
can survive. It is found experimentally [39] that the transfer 
function $<\widehat{S}({\bf u}_1)\widehat{S}^\ast({\bf u}_1 + {\bf u}_2) 
\widehat{S}({\bf u}_2)>$, is real and larger than zero 
up to the telescope cut off frequency. Due to the reality of this transfer 
function, the phase of the complex bispectra of the object is identical to the 
phase of the average bispectrum of the object specklegrams,
\begin{equation}
phase{\left<\widehat{I}({\bf u}_1,{\bf u}_2)\right>} = phase{\left<
\widehat{O}({\bf u}_1,{\bf u}_2)\right>}.
\end{equation}

\noindent
Therefore phase information about the object can be obtained directly from
the average bispectra without compensation of the transfer function. 
The measurement of the transfer function 
$<\widehat{S}({\bf u}_1)\widehat{S}^\ast({\bf u}_1 + {\bf u}_2) 
\widehat{S}({\bf u}_2)>$, may be calculated by evaluating specklegrams of 
an astronomical point source. Thus the object bispectrum is 
obtained from the equation (15), and the object triple correlation 
$O({\bf x}_1,{\bf x}_2)$ is obtained by 
inverse Fourier transforming of $\widehat{O}({\bf u}_1,{\bf u}_2)$. 

Another way is to evaluate the uncorrelated specklegram of the 
object (Gaussian speckle masking method), where no astronomical point source 
is essential to be measured. The advantage of this
method over previous one is that it can be performed in the correlation domain
as well as in Fourier domain. For a Gaussian model of the atmosphere [39]. 
\begin{eqnarray}
\label{gausm}
I({\bf x}_1,{\bf x}_2 = {\bf s}) - I_{nmm}({\bf x}_1,{\bf x}_2 = {\bf s}) 
- I_{nmn}({\bf x},{\bf x}_1 = {\bf s}) \nonumber \\
- I_{nnm}({\bf x}_1,{\bf x}_2 = {\bf s}) + 2I_{nmk}({\bf x}_1,{\bf x}_2 = 
{\bf s}) = O({\bf x}_1,{\bf x}_2 = {\bf s}).
\end{eqnarray}

\noindent
Here different indices in $I_{nmk}$ indicate that uncorrelated specklegrams 
are triple correlated. The equation $(\ref{gausm})$ can be used 
to calculate the triple correlation $O({\bf x}_1,{\bf x}_2 = {\bf s})$ for any 
masking vector ${\bf s}$. If ${\bf s}$ is selected suitably, a true image 
of the object can be obtained. In the case of complicated objects it is 
useful to choose a set of many different masking vectors in order to improve 
the signal to noise ratio. The information about suitable masking vectors is 
obtained from the object autocorrelation.

The image can be reconstructed by implementing recursive image reconstruction 
method in frequency domain, which is derived by assuming that a sampled 
version $O_{p,q}$ of the object bispectra is available,
\begin{equation}
\label{eq10}
O_{p,q} = O_p.O_q.O_{-p-q}  \quad {\rm  p,q=-N.....+N},
\end{equation}

\noindent
with ${\bf u}_1 = p.\Delta {\bf u}$, ${\bf u}_2 = q.\Delta {\bf u}$, 
and $P=-N......+N$, where $\Delta {\bf u}$ is 
a suitable sampling distance in the Fourier domain.
In this method, the modulus and phase of the complex object bispectrum $O_p$ are
calculated separately. Therefore $O_p$ is split in to:
\begin{equation}
\label{eq11}
O_p = |O_p| e^{(i\psi_p)}.
\end{equation}

\noindent
In the recursive image reconstruction method the modulus of the object 
Fourier transform is obtained as in speckle interferometry.  
Speckle interferometry data are produced by setting $p=0$ (or equivalently
by $q=0$ or $p=-q$) in $O_{p,q}$ i.e.,
\begin{equation}
\label{eq12}
O_{0.q} = O_0.O_q.O_{-q}=const. |O_q|^2,
\end{equation}

\noindent
where the fact that the spectrum of a real object is Hermitian i.e.,
$O_q = O_{-q}^\ast$ is used. By substituting the equation $(\ref{eq11})$ into 
$(\ref{eq10})$ the phase of the complex spectrum of the object is reconstructed,
\begin{equation}
|O_{p,q}| e^{(i \psi_{p+q})} = |O_p| e^{(i \psi_p)} \quad{\rm |O_q|} 
e^{(i \psi_q)} \quad{\rm |O_{-p-q}| e^{(i \psi_{-p-q})}}.
\end{equation}

\noindent
For $p+q=r$ and separate phase and modulus parts, the equation for phase will
be,
\begin{equation}
\label{eq13}
e^{(i \psi_r)} = e^{[i(\psi_{r-q}+\psi_q-\beta_{r-q,q})]},
\end{equation}

\noindent
where $\beta_{r-q,q}$ is the phase of 
\begin{equation}
O_{r-q,q} = |O_{r-q,q}| e^{(i \beta_{r-q,q})}.
\end{equation}
   
\begin{figure}
\centerline{\psfig{figure=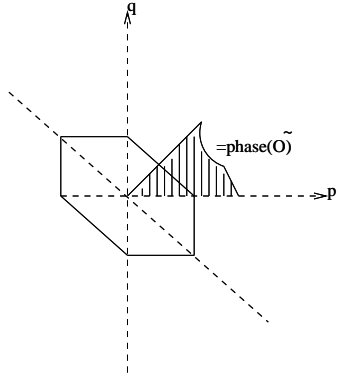,height=5cm}}
\caption{\fc Complex bispectrum $O_{p,q}$ of an object contains complete
information about the modulus and the phase of the object spectrum 
$|O_r|e^{(i\psi_r)}$. The modulus information $|O_r|$ can be reconstructed 
from one of the axes $p=0$, $q=0$ or $p=q$. The phase information 
$e^{(i\psi_r)}$ is contained in the area in between these distinguished axes.
Because of the eightfold symmetry of the bispectrum of a real function, only
one octant of the bispectrum contains non-redundant information, as indicated
by the area filled by lines.}
\end{figure}
\noindent
From this equation the phase factors $e^{(i\psi_r)}$ can be calculated
recursively. It is sufficient to calculate $O_r$ for positive $r$ since 
$O_r$ is Hermitian and therefore $\psi_r=\psi_{-r}$.
By setting $q=1$ in the equation $(\ref{eq13})$ yields
\begin{equation}	
\label{eq14}
e^{(i\psi_r)} = e^{[i(\psi_1+\psi_{r-1}-\beta_{r-1,1}]},
\end{equation}

\noindent
where
$\psi_0=\psi_1=0$, $r=2.....N$. $\psi_0$ is equal to zero because absolute
position in the reconstructed image is of no interest. In order to explain this 
let us start the algorithm with $r=0$. Both $\psi_0$ and $\beta_{0,1}$ are
zero due to the reality of $O({\bf x})$ and $O({\bf x}_1,{\bf x}_2)$. The 
recursive procedure is as follows,
\begin{eqnarray}
\psi_2 &=& 2\psi_1-\beta_{1,1} \\
\psi_3 &=& \psi_2-\psi_1-\beta_{2,1} \\
       &=& 3\psi_1-\beta_{1,1}-\beta_{2,1} \nonumber \\ 
& &................................... \nonumber\\
\psi_r &=& r\psi_1-\beta_{1,1}-\beta_{2,1}-.....\beta_{r-1,1}.
\end{eqnarray}

\noindent
Apparently, the Fourier phase $\psi_r$ and $O_r$ can be determined from the 
phases of the bispectrum, except for the linear term $r\psi_1$. This unknown
linear phase term corresponds to the unknown position of the object 
$O({\bf x}-{\bf x}_0)$.
For the reason explained above it is found that $\psi_1$ remains undeterminable.

The recursion given in the equation $(\ref{eq14})$ uses only the phase 
information
contained in a single line $(q=1)$ of the object bispectrum $O_{p,q}$.
Additional phase informations can be obtained by setting $q=2...N$ in the 
equation $(\ref{eq13})$. Therefore each phase $\psi_r$ has $(r-1)/2$ 
independent representations if $r$ is odd and $r/2$ representations
if $r$ is even. These different representations of the $\psi_r$ can be averaged.
To find the recursion formula, the equation $(\ref{eq13})$ is used which is 
insensitive to noise because of summation,
\begin{equation}
\label{eq15}
e^{(i\psi_r)} = const. \sum_{0<q\le r/2} e^{[i(\psi_q+\psi_{r-q} -
\beta_{r-q,q})]},
\end{equation}

\noindent
where $\psi_0=\psi_1=0$ and $r=2...N$ and the index $q$ in the summation was
selected such that all information contained in one octant of the object 
bispectrum is used to reconstruct the phases $\psi_r$. The remaining octants 
of the bispectrum do not supply additional information because of the inherent 
symmetry of bispectra as shown in Figure 7. By combining
the phase factors with the modulus of the object Fourier transform 
$(\ref{eq12})$, the recursive image is reconstructed. Inverse Fourier transform 
of the complex object spectrum yields the true image of the object.
\begin{figure}[h]
\centerline{\psfig{figure=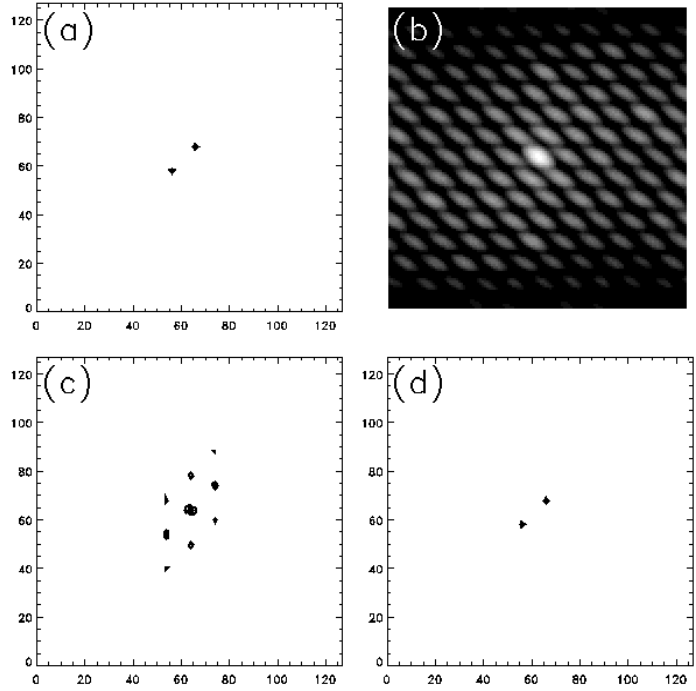,height=10cm}}
\caption{\fc Fourier phase of a simulated binary system: (a) 2-D maps of a 
simulated binary system, (b) 2-D representation of its 4-D bispectrum, (c) its 
triple correlation and (d) its reconstructed image.}
\end{figure}

Saha et al., [41] have developed 
a code based on the unit amplitude phasor recursive re-constructor. The
algorithm written in Interactive Data Language (IDL) takes about an hour for
processing 10 frames of size 128$\times$128 using the SPARC ULTRA workstation. 
The memory needed for the calculation
exceeds 160 MB if the array size is more than the said number. Since the
bispectrum is a 4-D function, it is difficult to represent it in
a 3-D co-ordinate system. Therefore, the
calculated values are stored in 1-D array and used them later to calculate the
phase by keeping track of the component frequencies [41].
Assuming $\psi(0, 0) = 0, \psi(0, \pm1) = 0$ and $\psi(\pm1, 0) = 0$, the 
phases are calculated by the unitary amplitude method. 
Though the memory required is independent of the
dimensionality of array, the time required to access an element in a 1-D 
array is much smaller than that in a 4-D array. In order to reduce the high 
frequency noise, Wiener filter parameter has been implemented in reconstruction 
stage. The bispectrum method has been tested 
with a computer simulated image by using a code developed by Saha et al. [41]. 
Figure 8 depicts (a) simulated binary system, (b) 2-D
representation of a 4-D bispectrum, (c) triple correlation and (d)
reconstructed image of the same binary system. 
\\

\noindent
{\s 6. Adaptive optics}      
\\

\noindent
The adaptive optics systems (AO) remove the turbulence induced wavefront 
distortions by introducing controllable counter wavefront distortion which 
both spatially and temporally follows that of the atmosphere. The purpose of 
this system is to (i) sense the wavefront perturbations, and (ii) compensate 
for them in real time [42, 43].  
\begin{figure}[h]
\centerline{\psfig{figure=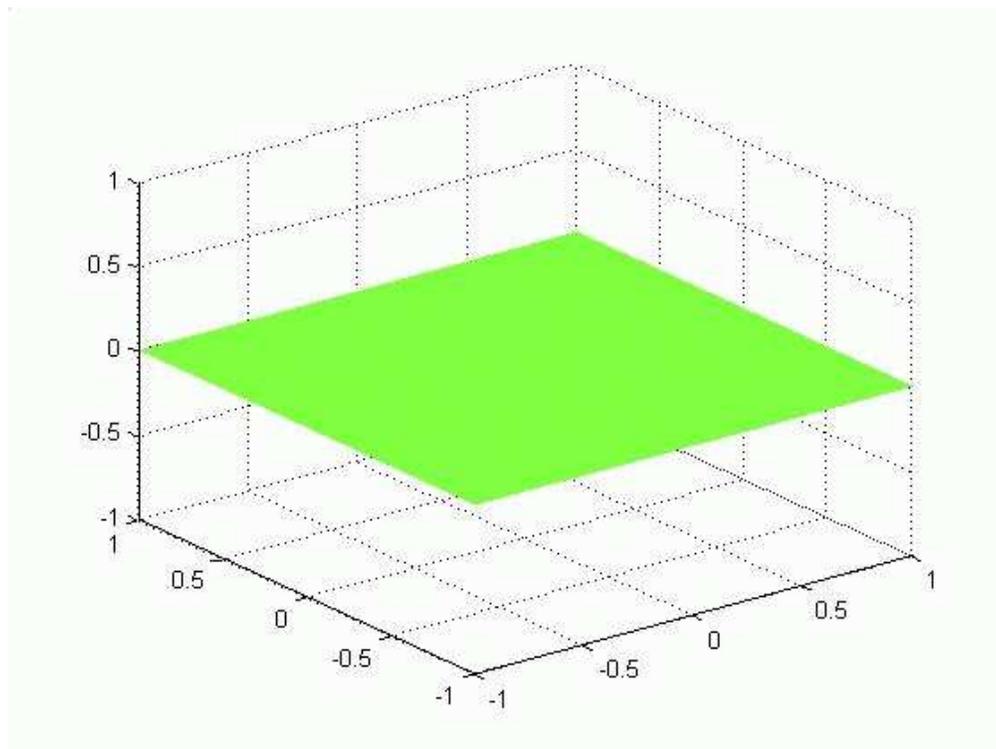,height=10cm}}
\caption{\fc Plane wavefront recorded at the laboratory (Courtesy: V. 
Chinnappan). 3-D picture is generated by using Matlab.}
\end{figure}

As the wind moves the eddies past the telescope aperture, the
tilt of the intercepted wavefront changes. Therefore, the wind
velocity dictates the speed. In order to take a corrective measure,
Greenwood [44] derives the mean square residual wavefront error as a
function of servoloop bandwidth for a first order controller, 
\begin{equation}
\sigma^2_{cl} = \left(\frac{{\it f}_G}{{\it f}_{3db}}\right)^{5/3} \, \,
{rad}^2,
\end{equation}

\noindent
where ${\it f}_{3db}$ is the closed loop bandwidth of the wavefront compensator
and ${\it f}_G = 0.426{\it v}/r_0$ the Greenwood frequency, in which ${\it v}$ 
is the wind velocity in the turbulent layer of air. 
\begin{figure}[h]
\centerline{\psfig{figure=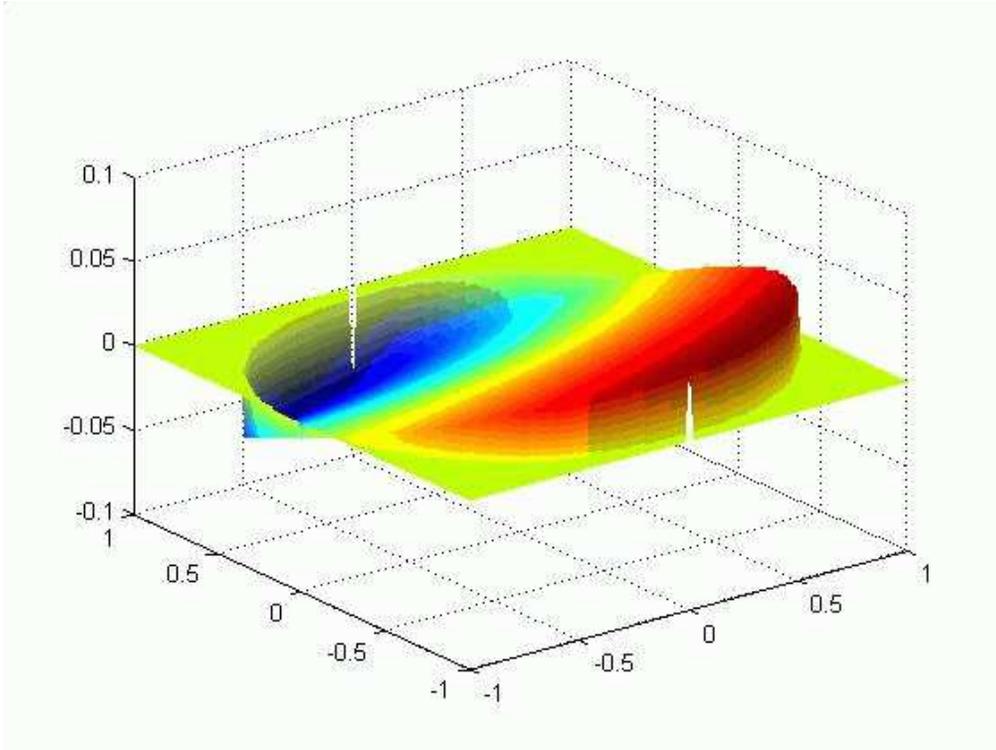,height=10cm}}
\caption{\fc Laboratory measurement of wavefront tilt (Courtesy: V. 
Chinnappan).}
\end{figure}

Figure (9) depicts the plane wavefront that is generated at the 
laboratory with a laser source offering zero volt to the tip-tilt mirror. 
While Figure (10) depicts the wavefront tilt measured with the same source
after applying one volt to the said mirror [45]. These images are grabbed by a 
CMOS imager based Shack-Hartman (SH) sensor. These plane 
and tilted wavefronts resemble to the wavefronts arriving to a detector 
from a distant star before and after passing through the turbulence of the 
atmosphere respectively. The laboratory experiment shows only tilt as a
major error, while in the case of atmosphere, the wavefront will
have complicated contours. Nevertheless, a reverse situation can be created by 
employing the AO systems in order to improve the throughput of the large 
telescope. 

The required components for implementing an AO system are: (i) wavefront sensor,
(ii) wavefront phase error computation and (iii) a flexible mirror whose surface
is electronically controlled in real time to create a conjugate surface enabling
to compensate the wavefront distortion [43, 46]. In order to remove the low 
frequency tilt error, generally the incoming collimated beam is fed by a 
tip-tilt mirror. After traveling further, it reflects off of a deformable 
mirror (DM) that eliminates high frequency wavefront errors (see 
Figure 11). A beam-splitter divides the beam into two parts: one is
directed to the wavefront sensor to measure the residual error in the wavefront
and to provide information to the actuator control computer to compute the DM
actuator voltages and the other is focused to form an image. The control 
system acts as feedback loop; the next cycle corrects the small errors of the
last cycle. This process should be at speeds commensurate with the rate of 
change of the corrugated wave-front phase errors. 
\begin{figure}[h]
\centerline{\psfig{figure=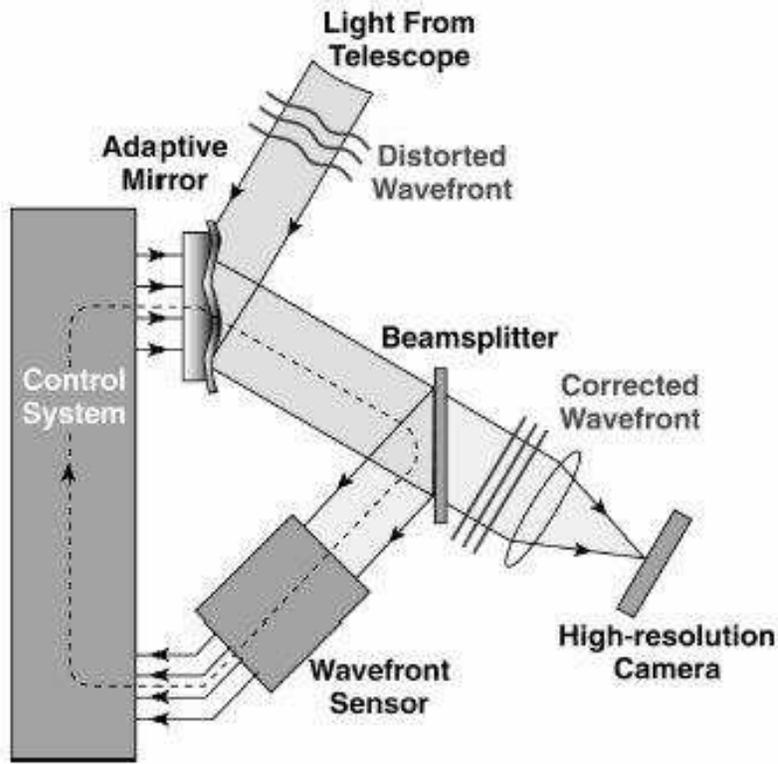,height=10.5cm}}
\caption{\fc Schematic diagram of the adaptive optics system.}
\end{figure}

The shape of the image compensation devices is controlled such that perturbed
incident wavefront phase-shifts are canceled as the optical field bounces
from their surfaces. It is easier to achieve
diffraction-limited information using AO systems at longer wavelengths since
$r_0 \propto \lambda^{6/5}$, which implies that the width of seeing limited 
images, $1.22\lambda/r_0 \propto \lambda^{-1/5}$ varies with $\lambda$.
The number of degrees of freedom, i.e., the number of actuators on the
DM and the number of sub-aperture in the wavefront sensor,
in an AO system is determined by,
\begin{equation}
(D/r_0)^2 \propto \lambda^{-12/5}.
\end{equation}

Another way to correct the disturbance in real time is usage of adaptive 
secondary mirror (ASM) that makes relay optics obsolete [47].
The other notable advantages are: (i) enhanced photon throughput that
measures the proportion of light which is transmitted through an optical
set-up, (ii) introduction of negligible extra IR emissivity, (iii) causes no
extra polarization, and (iv) non-addition of reflective losses [48].
Due to the interactuator spacing, the resonant frequency of such a
mirror may be lower than the AO bandwidth. The ASM system uses a SH sensor with
an array of small lenslets, which adds two extra refractive surfaces to the
wavefront sensor optical beam [49]. An $f/15$ AO secondary with
336 actuators is installed on the 6.5~m Telescope of MMT observatory, Mt.
Hopkins, Arizona. Figure 12 depicts the real time image of ADS 1585 [50]
with a resolution of 0.07$^{\prime\prime}$ (FWHM). This image is acquired with 
the 6.5~m MMT adaptive secondary mirror at H Band (1.65~$\mu$m). 
\begin{figure}[h]
\centerline{\psfig{figure=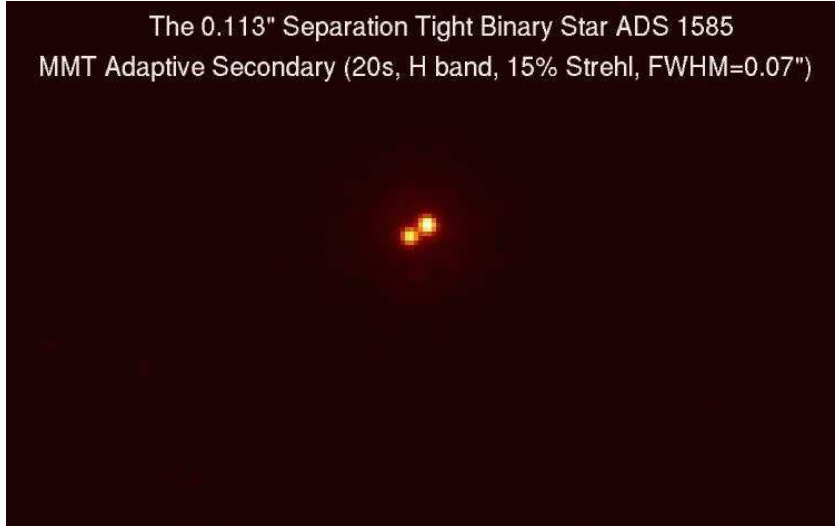,height=7cm}}
\caption{\fc Real time image of ADS 1585 [50](Courtesy: L. Close).}
\end{figure}
\\

\noindent
{\s 7. Summary}      
\\

\noindent
Speckle interferometry is a boon for the observational astronomy and has been
contributing to the study of a variety of astrophysical problems.
Among others the most important observations made by such technique 
is the discovery of compact cluster, R136a (HD~38268), of 
Doradus nebula in the Large Magellanic Clouds [14]. The
object was thought to be the most massive star with a solar mass of 
$\sim$~2500M$_\odot$. In one of the IAU conferences, one full day was 
spent in discussing about its probability of a black hole [51].
The reconstructed image with speckle masking 
technique depicted more than 40 components in the 
4.9$^{\prime\prime}\times 4.9^{\prime\prime}$ field; the closest binaries were 
found to be 0.03$^{\prime\prime}$ and 0.05$^{\prime\prime}$ [52].
Recent observations with adaptive optics system by Brandl et al. [53] 
have revealed over 500 stars within the field of view 
12.8$^{\prime\prime}\times 12.8^{\prime\prime}$ covering a magnitude range 
of 11.2. 

Studies of binary stars play a fundamental role in measuring stellar masses. 
Mass determinations of stars provide a benchmark for stellar evolution 
calculations. A long-term benefit of speckle imaging is a better calibration 
of the main-sequence mass-luminosity relationship.
Speckle interferometric technique has made major inroads into decreasing the 
gap between visual and spectroscopic binaries by achieving angular resolution 
down to 20 milliarcsec (mas). Most measurements have been made at large aperture
telescopes by groups in France, Russia, and the United States. Programmes of
binary star interferometry are being carried out at telescopes of moderate and
small aperture too. But measurements from the southern hemisphere continue to be
rare. Many rapidly moving southern binaries are being ignored and a number of
discoveries are yet to be confirmed. 
Most of the late-type stars are available in the vicinity of sun. All
known stars, within 5 pc radius from the sun are red dwarfs with m$_v >$ +15.
Due to the intrinsically faint nature of K- and M- dwarfs, their physical
properties are not studied extensively. These dwarfs may often be close
binaries which can also be detected. Figure 13 depicts the blind iterative 
deconvolution (BID) technique that applies to imaging in
general which covers methods spanning from simple linear
deconvolution algorithms to complex non-linear algorithms,
reconstruction of a binary system, HR~5138 [54], in which the separation 
between the components is found to be about 0.27$^{\prime\prime}$.
\begin{figure}[h]
\centerline{\psfig{figure=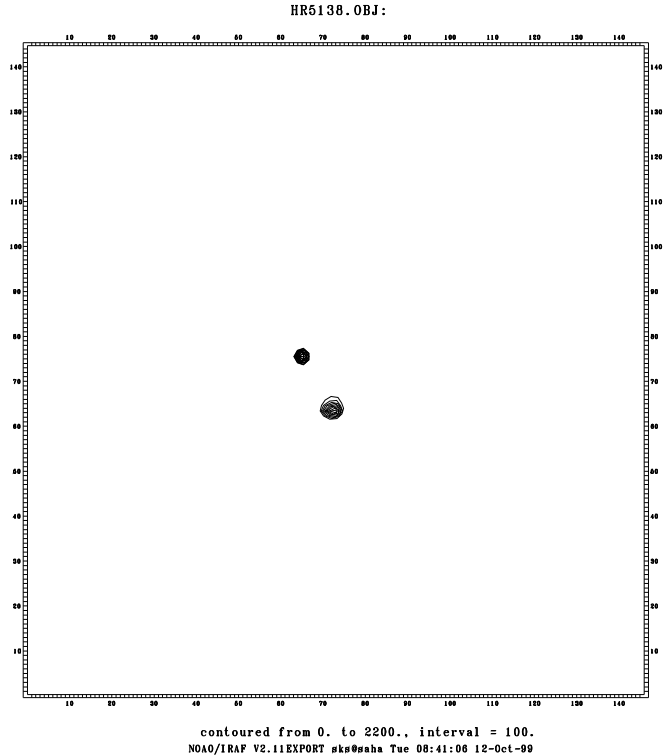,height=10cm,angle=270}}
\caption{\fc The blind iterative deconvolution (BID) reconstruction of a 
binary system, HR~5138 [54].}
\end{figure}

Another important field of observational astronomy is
the study of the physical processes, viz., temperature, density and velocity
of gas in the active regions of the active galactic nuclei (AGN); optical
imaging in the light of emission lines on the sub-arcsecond scales can reveal
the structure of the narrowline region. Binary black holes have been suggested
to be the cause of the two components observed in the profiles of
the broad emission lines of Arp 102 and a few other AGNs [55].

Another field of interest is the study of Quasars (QSO) that may be 
gravitationally lensed by stellar objects, viz., stars, galaxies, cluster of
galaxies etc. located in the line of sight. The aim of the high angular
imagery of those QSO's is to find their structure and components as well as
to determine their number and structure as a probe of the distribution of the 
mass in the Universe. 

The diffraction limited phase retrieval of a degraded image is an important
subject that is being implemented in other branches of physics too, e.g., 
electron microscopy, wavefront sensing, and crystallography. A second-order 
moment (power spectrum) analysis provides only the modulus of the object FT, 
whereas a third-order moment (bispectrum) analysis yields the phase allowing 
the object to be fully reconstructed. A more recent attempt to go beyond the 
third order, e.g., fourth-order moment (trispectrum), illustrates its utility 
in finding optimal quadratic statistics through the weak gravitational lensing 
effect [56], however, its implementation in optical imaging is a difficult 
computational task.

The programme of speckle imaging at VBT
has been a successful one. Now we are in a position to obtain informations of
Fourier phase of the objects too [41]. Mapping of the certain 
interesting objects, viz., active galactic nuclei, proto planetary nebulae 
will be undertaken in near future. However developing a school in the
field of interferometry in optical/IR band in India is an important task to cope
with the present day technological advancement across the globe. 
Several long baseline optical interferometers are in operation and made an
impact in astrophysical studies [3, 4]. Work on a few interferometric
projects by employing large telescopes are in progress. One such project called 
Large Binocular Telescope (LBT), in which the mirrors are co-mounted on a 
fully steerable alt-az mount, offers unprecedented spatial resolution of the 
order of 8-9~mas at $\lambda \sim 1~\mu$m [57]; the information
in the $u, v$ plane can be continuously combined or coadded. 
Developing a long baseline interferometer using such 
a technology will have far reaching impact on astrophysics, thus offering the
possibilities for direct measurement of all the basic physical parameters
for a large number of stars. With instruments as powerful as the
current generation of working or planned interferometers, the element
of serendipity will bring many surprises to astronomy.
\\

\noindent
{\s Acknowledgement}: I express my gratitude to Dr. L. Close for providing
the real time image of ADS 1585.      
}
\\

\noindent
{\s References}      
\\

{\fc
\noindent
1. Fried D C, {\it J. Opt. Soc. Am.}, 56 (1966), 1972. 

\noindent
2. Labeyrie A, {\it Astron. Astrophys.}, 6 (1970), 85.

\noindent
3. Saha S K, {\it Bull. Astron. Soc. Ind.}, 27 (1999), 443.

\noindent
4. Saha S K, {\it Rev. Mod. Phys.}, 74 (2002), 551.

\noindent
5. Hartkopf W I, McAlister H A, Mason B D, CHARA Contrib. No. 4,
{\it Third Catalog of Interferometric Measurements of Binary Stars}, (1997) W.I.

\noindent
6. Bonneau D, Labeyrie A, {\it Astrophys. J}, 181 (1973), L1.

\noindent
7. Labeyrie A, Koechlin L, Bonneau D, Blazit A, Foy R, 
{\it Astrophys. J}, 218 (1977), L75.

\noindent
8. Bonneau D, Foy R, {\it Astron. Astrophys.}, 92 (1980), L1.

\noindent
9. Drummond J, Eckart A, Hege E, {\it Icarus}, 73 (1988), 1.

\noindent
10. Saha S K, Rajamohan, Rao P V, Som Sundar G, Swaminathan R, Lokanadham B,
{\it Bull. Astron. Soc. Ind.}, 25 (1997), 563.

\noindent
11. Wood P, Bessel M, Dopita M, {\it Astrophys. J}, 311 (1986), 632.

\noindent
12. Osterbart R, Balega Y, Weigelt G, Langer N, {\it Proc. IAU symp. 180},
eds., H Habing \& G Lamers, (1996), 362.

\noindent
13. Ebstein S, Carleton N P, Papaliolios C, {\it Astrophys. J}, 336 (1989), 103.

\noindent
14. Weigelt G P, Baier G, {\it Astron. Astrophys.}, 150 (1985), L18.

\noindent
15. Foy R, Bonneau D, Blazit A, {\it Astron. Astrophys.}, 149 (1985), L13.

\noindent
16. Baba N, Kuwamura S, Miura N, Norimoto Y, {\it Astrophys. J.}, 431 (1994),
L111.

\noindent
17. Babcock H W, {\it Pub. Astron. Soc. Pac.}, 65 (1953), 229.
 
\noindent
18. Rousset G, Fontanella J C, Kem P, Gigan P, Rigaut F, L\'ena P, Boyer P,
Jagourel P, Gaffard J P, Merkle F, {\it Astron. Astrophys.}, 230 (1990),
L29.

\noindent
19. Liang J, Williams D R, Millar D T, {\it J. Opt. Soc. Am. A}, 14 (1997), 
2884.
 
\noindent
20. Labeyrie A, {\it Astrophys. J.}, 196 (1975), L71.

\noindent
21. Mourard D, Bosc I, Labeyrie A, Koechlin L, Saha S, {\it Nature},
342 (1989), 520.

\noindent
22. Sirohi R S, editor, {\it Selected papers in Speckle Metrology}, (1991), 
Milestone Series MS 35, Bellingham, Washington, SPIE, Optical Engineering Press.

\noindent
23. Sirohi R S, editor, {\it Speckle Metrology}, (1993), Dekkar, N Y.

\noindent
24. Sirohi R S, {\it Contemporary Physics}, 43 (2002), 161.

\noindent
25. Saha S K, {\it Ind. J. Phys.}, 73B (1999), 552.

\noindent
26. Kolmogorov A, {\it Turbulence}, eds. S K Friedlander \& L Topper, (1961),
151.

\noindent
27. Tatarski V I, {\it Wave Propagation in Turbulent Medium}, (1967) Dover, N Y.

\noindent
28. Ishimaru A, {\it Wave Propagation and Scattering in Random Media},
(1978), Academic Press, N Y.

\noindent
29. Saha S K, Chinnappan V, {\it Bull. Astron. Soc. Ind.}, 27 (1999), 327.

\noindent
30. Labeyrie A, {\it 15th. Advanced Course, Swiss Society of Astrophys. 
and Astron.} ed.. A Benz, M Huber and M. Mayor, (1985), 170.

\noindent
31. Born M, Wolf E, {\it Principles of Optics}, (1984), Pergamon Press.

\noindent
32. Saha S K, Jayarajan A P, Sudheendra G, Umesh Chandra A,  
{\it Bull. Astron. Soc. Ind.}, 25 (1997), 379.

\noindent
33. Saha S K, Sudheendra G, Umesh Chandra A, Chinnappan V, {\it Experimental 
Astronomy}, 9 (1999), 39. 

\noindent
34. Saha S K, Proc. {\it Young Astrophysicists of To-day's India}, (2001),
http://xxx.lanl.gov/astro-ph/0003125. 

\noindent
35. Foy R, {\it Instrumentation for Ground Based Optical Astronomy - 
Present and Future}, (1988), ed. L. Robinson, Springer-Verlag, N Y, 345.

\noindent
36. Saha S K, Maitra D, {\it Ind. J. Phys.}, 75B (2001), 391.

\noindent
37. Chinnappan V, Saha S K, and Faseehana, {\it Kod. Obs. Bull.} 11 
(1991), 87.

\noindent
38. Weigelt G P, {\it Opt Communication}, 21 (1977), 55. 

\noindent
39. Lohmann A W, Weigelt G P, Wirnitzer B, {\it Applied Optics} 22 (1983), 
4028. 

\noindent
40. Reinheimer T, Weigelt G P, {\it Astron. Astrophys.} 176 (1987), L17. 

\noindent
41. Saha S K, Sridharan R, Sankarasubramanian K, {\it Speckle Image 
Reconstruction of Binary Stars}, (1999), Presented at the  
XIX Astron. Soc. Ind. conf, Bangalore.

\noindent
42. Beckers J M, {\it Annual Rev. Astron. Astrophys.} 31 (1993), 13. 

\noindent
43. Roggemann M C, Welsh B M, Fugate R Q, {\it Rev. Modern Phys.}, 69 (1997), 
437.

\noindent
44. Greenwood D P, {\it J. Opt. Soc. Am.}, 67 (1977), 390.

\noindent
45. Chinnappan V, {\it Private Communication}, (2003).

\noindent
46. Roddier F (editor), {\it Adaptive Optics in Astronomy}, (1999), Cambridge
Univ. Press.

\noindent
47. Bruns D, Barnett T, Sandler D, {\it SPIE.}, 2871 (1997), 890.

\noindent
48. Lee J, Bigelow B, Walker D, Doel A, Bingham R, {\it Pub. Astron. Soc.
Pac}, 112 (2000), 97.

\noindent
49. Lloyd-Hart M, 2000, {\it Pub. Astron. Soc. Pac}, 112 (2000), 264.

\noindent
50. Close L, http://athene.as.arizona.edu/~lclose/AOPRESS/, (2003).

\noindent
51. Cassinelli J P, Mathis J C, Savage B R, {\it Science}, 212 (1911), 1497.
 
\noindent
52. Pehlemann E, Hofmann K -H, Weigelt G P, {\it Astron. Astrophys.},
256 (1992), 701.
 
\noindent
53. Brandl B, Sams B J, Bertoldi F, Eckart A, Genzel R, Drapatz S, Hofmann R, 
Lowe M, Quirrenbach A, {\it Astrophys. J.}, 466 (1996), 254.  

\noindent
54. Saha S K, Venkatakrishnan P V, {\it Bull. Astron. Soc. Ind.}, 25 (1997), 
329.

\noindent
55. Ulrich M. -H., Proc. ESO-NOAO conf. {\it High Resolution Imaging
Interferometry}, ed. F Merkle, Garching bei M\"unchen, FRG, (1988), 33.

\noindent
56. Hu W, http://xxx.lanl.gov/astro-ph/0105117, (2001).

\noindent
57. Weinger P, {\it Private Communication}, (2001).

\end{document}